\documentclass[12pt,preprintnumbers]{article}
\usepackage{amsmath, amssymb, wasysym, slashed, multirow,hyperref}
\usepackage{pdfpages}
\usepackage{cite}
\usepackage{dsfont}
\usepackage{times}
\usepackage{xcolor}
\usepackage{graphicx,color}  %
\usepackage{bm}  %
\usepackage{enumitem}
\usepackage{braket}
\newenvironment{sciabstract}{%
\begin{quote} }
{\end{quote}}

\newcounter{lastnote}

\usepackage{fancyhdr}
\fancypagestyle{plain}{%
	\fancyhead[R]{}

}

\title{Minimal Proper-time in  Quantum Field Theory }

\author
{Alessio Maiezza$^{1,2\ast}$, Juan Carlos Vasquez$^{3\dagger}$\\
\\
\normalsize{$^{1}$Dipartimento di Scienze Fisiche e Chimiche, Universit\`a} \\ \\
\normalsize{degli Studi dell'Aquila, via Vetoio, I-67100, L'Aquila, Italy,}\\ \\
\normalsize{$^{2}$INFN, Laboratori Nazionali del Gran Sasso, 67010 Assergi, L'Aquila, Italy,}\\ \\
\normalsize{$^{3}$SISSA, Via Bonomea 265, I-34136 Trieste, Italy} \\
\\
\small{ E-mail: alessiomaiezza@gmail.com$^{\ast}$, juancarlos8866@gmail.com$^{\dagger}$}
}

\date{}

\begin{document}


\baselineskip16pt 


\maketitle


\begin{sciabstract}
We propose a generalization of quantum field theory within Schrödinger’s functional representation, inspired by Nambu’s proper-time formulation of quantum mechanics.  The key motivation for this generalization is to incorporate a fundamental, Lorentz-invariant minimum scale, which in this formulation is played by a minimal proper time $\tau_{\min}$. The introduction of $\tau_{\min}$ leads to several significant effects at very high energies: it modifies the Heisenberg uncertainty principle, induces a controlled violation of unitarity, and suppresses high-energy modes. This minimal scale renders the theory asymptotically safe through a mechanism akin to dimensional reduction, while reproducing all the standard results at low energies, where quantum field theory  emerges. Remarkably, the same framework can accommodate a deterministic regime at energies approaching the Planck scale. These features suggest that a minimal proper-time formulation renders the quantum field theory an effective but finite theory, superseded at trans-Planckian energies. 
\end{sciabstract}

\section{Introduction}

Quantum Field Theory (QFT) in the Schrödinger representation provides a powerful framework. The main dynamical object is the wave functional that depends on the field configurations. Unlike the Heisenberg picture, this approach makes transparent the analogy with quantum mechanics. 

In this work, we develop a new formulation of QFT in the Schrödinger representation inspired by Nambu’s proper-time approach to quantum mechanics (QM) \cite{Nambu:1950rs}. Nambu questioned the special status of the time coordinate  in quantum theory and proposed that the evolution of quantum systems should be governed by an internal, Lorentz-invariant parameter, namely, the proper time\footnote{The approach can be traced back to the work of Feynman \cite{PhysRev.76.749} and extensively developed by Nambu in a form on which we build on in this article. Originally, the idea of a proper time formulation was formulated by Fock \cite{Fock:1937dy} within relativistic dynamics of a particle.}. As we shall see, the time evolution is governed not by the Hamiltonian,  but by an evolution equation involving the proper time that automatically  leads to the standard Schrödinger equation. This conceptual shift enables all spacetime coordinates to be treated on an equal footing, consistent with special relativity.   This approach is inspired by relativistic mechanics, where particle dynamics is often parameterized by proper time, and it aims to build a covariant quantum theory more naturally aligned with the structure of spacetime. From a four-dimensional perspective, time no longer indicates progress but is just another dimension. Creation and annihilation are simply changes in a particle's direction through time —  one particle from past to future and the other  from the future to past. In this view, virtual pair-creation is a single-particle loop in  space-time, and real particles are such that their orbits are not closed and reach infinity.

The key conceptual contribution of this work is the interpretation of the fundamental Planck scale as  the inverse of the minimum proper time $\tau_{\min}>0$, which is automatically the same in all inertial frames.  This may be seen as
the other side of the medal of the minimal length scenario, often proposed in the literature \cite{Mead:1964zz,GARAY_1995,Kempf_1995,Padmanabhan:1996ap,Modesto_2009,Nicolini_2011,Hossenfelder_2013,Bosso_2023,Bosso_2024}. One motivation for considering a minimal length in flat space comes from general arguments from black hole physics and quantum mechanics \cite{Maggiore:1993rv,Aurilia:2013psa}.

The resulting framework has several interesting implications. \emph{Suppressed non-unitary evolution:} The presence of $\tau_{\min}$ introduces a controlled, physically motivated non-unitarity in the evolution of wave functionals. This kind of controlled, effective non‑unitarity resonates with discussions in the black hole information paradox literature at the semiclassical level -- see, for a review, \cite{Harlow_2016}.

\emph{Modified canonical commutation relations:} the previous point leads to a deformation of canonical commutation relations and an effective running Planck constant at high energies, thereby corresponding to a modification of Heisenberg's uncertainty relation.

\emph{Schwinger’s propagator:} the proper‑time Schrödinger framework naturally connects to the Schwinger proper‑time representation of the propagator. When a minimal proper time $\tau_{min}$ is postulated, the Schwinger parameter is endowed with a physical lower bound, yielding an UV suppression of loop integrals.

\emph{UV finiteness:} a direct consequence is that the minimal proper-time regularizes loop integrals, rendering the theory finite at all  orders in perturbation theory. The cutoff $\tau_{\min}$ acts as a natural UV regulator, in the spirit of Schwinger’s method but with a physical interpretation. The exponential suppression at high energies can be reinterpreted as a form of dimensional reduction in the deep UV, with automatic asymptotic safety.

The rest of this paper is organized as follows. In sec. \ref{sec-nambu}, we highlight the main feature of the seminal work of Nambu \cite{Nambu:1950rs}. 
In sec. \ref{sec-schoed}, we recall the Schrödinger representation in QFT, specifically for a scalar real field, since it is not commonly used. Then, the succeeding section, \ref{sec-nambu-QFT}, merges the concepts in the previous sections, namely, importing Nambu's approach into QFT. With this formalism at hand, we introduce a minimal proper time, in the main section \ref{sec-constraint}, as a fundamental physical scale. All the consequences of this scale are examined in sec. \ref{sec-emergent}. Finally, we offer our conclusions in sec. \ref{sec-end}. The article is completed by some appendices containing technical details: in the appendix \ref{appedix:inner_products}, we summarize the inner products in several formalism; in \ref{appendix:interpretation}, we elaborate on the free-particle physical picture in the light of the proper-time formalisms; in \ref{appendix:RGE}, we revisit renormalization group in the proper-time language.

\section{Overview on Nambu's Proper-Time Quantum Mechanics}\label{sec-nambu}

The central object of the theory is the wavefunction $\psi(x, \tau)$ (denoting $x=(t,\vec{x})$), now depending on both the spatial coordinates and the auxiliary proper time $\tau$. The evolution equation for a particle takes the form:
\begin{equation} \label{base}
i\hbar \frac{\partial}{\partial \tau} \psi(x, \tau) = \hat{H}' \psi(x, \tau),
\end{equation}
where the operator $\hat{H}'$ is defined as:
\begin{equation}
\hat{H}' = \hat{H} - i\hbar \frac{\partial}{\partial t}.
\end{equation}
Here, $\hat{H}$ is the standard Hamiltonian:
\begin{equation}
\hat{H} = -\frac{\hbar^2}{2m} \nabla^2 + V(x),
\end{equation}
and $t$ is still treated as a parameter inside the wavefunction and the potential, but no longer as the evolution parameter.

By transforming \eqref{base} into an eigenfunction problem
\begin{equation}
 \psi(x,\tau) = \psi(x) e^{-i\,\lambda\, \tau}   
\end{equation}
one has
\begin{equation}
 \hat{H'}\psi(x) = \lambda \psi(x) 
\end{equation}
The eigenfunction associated with the physical $\lambda=0$ sector can be extracted by integrating over the proper time $\tau$. Mathematically, this half-line integration yields a complex distribution,
\begin{equation}
 \int_0^{\infty} d\tau\, \psi(x,\tau) = \int_0^{\infty} d\tau\, \psi(x) e^{-i\,\lambda\, \tau} = \left[ \pi \delta(\lambda) - i \mathcal{P}\left(\frac{1}{\lambda}\right) \right] \psi(x)\,,
\end{equation}
where $\delta$ is the Dirac delta function and $\mathcal{P}$ denotes the Cauchy principal value. As we will discuss in more detail later (see Eq. 53), it is the real part of this integration that properly enforces the $\lambda=0$ constraint, projecting the extended state onto the standard physical wave function. We can thus formally define the recovered physical state $\psi^R(x)$ by isolating this delta-function contribution:
\begin{equation}\label{Re_delta}
 \psi^R(x) \propto \text{Re}\left( \int_0^{\infty} d\tau\, \psi(x,\tau) \right) = \pi \delta(\lambda) \psi(x)\,.
\end{equation}
Thus, the physical states are constrained by the condition:
\begin{equation} \label{eq:Schrödinger_real}
\hat{H}' \psi^R(x) = -i\,\delta(\lambda)\, \lambda \,\psi^R(x) = 0
\end{equation}
which imposes a reparametrization invariance under shifts of $\tau$. This constraint plays a role analogous to the Hamiltonian constraint in generally covariant systems such as the Wheeler-DeWitt equation. When the constraint is satisfied exactly, the formalism reduces to standard quantum mechanics. The conventional Schrödinger equation is recovered by solving the constraint:
\begin{equation}
\hat{H} \psi^R = i\hbar \frac{\partial}{\partial t} \psi^R,
\end{equation}
thus recovering the usual time evolution in $t$.

This reformulation frames quantum mechanics in a covariant form, allowing $t$ to be treated as a  label rather than a fundamental parameter of evolution. However, the theory presents troubles, among others, when generalizing it to many-body systems -- e.g, see \cite{Bohm2006-BOHTUU-5}, p. 238. This suggests that, as usual, one has to resort to a QFT description in place of particle mechanics. Yet, the Nambu formalism turns out to be illuminating also in QFT, motivating what follows.

\section{Highlighting scalar QFT in Schrödinger representation}\label{sec-schoed}

The Schrödinger representation  provides the most natural framework to generalize QFT to incorporate Nambu's proper-time formalism  with a functional Schrödinger  equation, in place of the usual Schrödinger differential equation. 

Since the Schrödinger representation is uncommon in QFT, we summarize its main features using a scalar field as a prototype (cf. \cite{Hatfield:1992rz}, p. 200).

In analogy with QM, where one has,
\begin{equation}
\hat{x}_i \lvert x_i \rangle = x_i \, \lvert x_i \rangle\,
\end{equation}
where, $i=1,2,3$. In QFT, the notion of coordinate $x$ is replaced with that of the field:
\begin{equation}
\hat{\phi}(\vec{x}) \lvert \phi \rangle = \phi(\vec{x}) \, \lvert \phi \rangle.
\end{equation}
Note that the hatted symbol indicates the operator, while the plain symbol denotes its eigenvalues, which is a function for the case at hand.

In the Shrodinger representation of QFT, the wave-function $\psi(\vec{x},t)$ is  replaced by the wave-functional,
\begin{equation}
\Psi[\phi, t],
\end{equation}
obeying the  functional Schrödinger equation:
\begin{equation}\label{Seq1}
\hat{H} \Psi[\phi, t] = i \frac{\partial}{\partial t} \Psi[\phi, t].
\end{equation}
Consequently, the framework of Quantum Field Theory (QFT) employed in the Schrödinger representation is rooted in functional analysis. For instance,
\begin{equation}
\frac{\delta \phi(\vec{x})}{\delta \phi(\vec{y})} = \delta(\vec{x}-\vec{y}) = \left[\frac{\delta}{\delta \phi(\vec{y})}, \phi(\vec{y}) \right],
\end{equation}
where $\frac{\delta}{\delta \phi(\vec{x})}$ denotes the functional derivative. Subsequently, by aligning the aforementioned commutator with the canonical commutation relations (CCR) pertaining to the free field, it is concluded that
\begin{align}
\text{operator formalism} & \longleftrightarrow  \text{functional formalism} \\
\hat{\pi}=-i \hbar\dot{\hat{\phi}}\hspace{2em} & \longleftrightarrow \hspace{2em}\pi=- i \hbar \frac{\delta}{\delta \phi(\vec{x})},
\end{align}
where $\hat{\pi}$ or $\pi$ denote the conjugate field in the two languages. 

The free Hamiltonian $\hat{H}$ for a real scalar free field reads,
\begin{equation}\label{Seq2}
\hat{H} = \int d^3x \left[ -\frac{\hbar^2}{2} \frac{\delta^2}{\delta \phi(\vec{x})^2} + \frac{1}{2} \left( \nabla \phi(\vec{x})^2  + m^2 \phi(\vec{x})^2 \right) \right],
\end{equation}
defining \eqref{Seq1}.

Notice the limitation of \eqref{Seq2} to ordinary quantum mechanics. When we transition from the infinite-dimensional field variable $\phi(\vec{x})$ to a one-dimensional variable, $q$, the gradient and integral in \eqref{Seq2} vanish because there’s no longer a space ($\vec{x}$) on which we can integrate or differentiate. Consequently, the second functional derivative transforms into a regular derivative, $d^2/d q^2$. As a result, \eqref{Seq2} simplifies to the ordinary Hamiltonian of a one-dimensional Harmonic oscillator with frequency $m$, as it should.

\subsection{Free Propagator}

In this section, we demonstrate how to express the two-point Green function using the Schrödinger representation functional formalism. To achieve this, we write the two-point function in the Schrödinger representation as follows:  
\begin{equation}
\langle 0 \vert \hat{\phi}(x') \hat{\phi}(x) \vert 0 \rangle \rightarrow \langle 0,t' \vert \hat{\phi}(\vec{x}') \hat{\phi}(\vec{x}) \vert 0,t \rangle.
\end{equation}
Let us set $t=0$ and rename $t'=t$.

Next, we use the fact that:
\begin{equation}\label{appl}
\hat{\phi}(\vec{x}) \lvert \phi \rangle = \phi(\vec{x}) \lvert \phi \rangle \quad \Rightarrow \quad \langle \phi \vert \hat{\phi}(\vec{x}) = \phi(\vec{x}) \langle \phi \vert.
\end{equation}
We now insert the identity operator via the completeness relation:
\begin{equation}
1 = \int \mathcal{D}\phi\, \lvert \phi \rangle \langle \phi \rvert,
\end{equation}
between the two field operators:
\begin{align}
\langle 0 \vert \hat{\phi}(x') \hat{\phi}(x) \vert 0 \rangle
&= \langle 0 \vert \hat{\phi}(x') \left( \int \mathcal{D}\phi\, \lvert \phi \rangle \langle \phi \rvert \right) \hat{\phi}(x) \vert 0 \rangle \nonumber \\
&= \int \mathcal{D}\phi\, \langle 0 \vert \hat{\phi}(x') \lvert \phi \rangle \langle \phi \vert \hat{\phi}(x) \vert 0 \rangle.
\end{align}
Using \eqref{appl} leads to
\begin{equation}
\langle 0,t'=t \vert \hat{\phi}(\vec{x}') \hat{\phi}(\vec{x}) \vert 0 \rangle = \int \mathcal{D}\phi\, \phi(\vec{x})\, \phi(\vec{x}')\, \langle 0 \vert \phi \rangle \langle \phi \vert 0,t \rangle.
\end{equation}
Next, one has to recognize the wavefunctional ground state.
To this aim, leveraging the completeness of the base $\vert \phi \rangle$, we define,
\begin{equation}
\vert 0 \rangle := \int \mathcal{D} \phi' \Psi_0[\phi'] \vert \phi' \rangle
\end{equation}
or
\begin{equation}
\langle \phi \vert 0 \rangle = \int \mathcal{D} \phi' \Psi_0[\phi'] \langle \phi \vert \phi' \rangle = \int \mathcal{D} \phi' \Psi_0[\phi'] \delta[\phi-\phi']= \Psi_0[\phi].
\end{equation}
Therefore, one has for any $t$,
\begin{equation}
\langle \phi \vert 0, t \rangle = \Psi_0[\phi, t],
\end{equation}
thus we arrive at
\begin{equation}
\langle 0 \vert \hat{\phi}(x') \hat{\phi}(x) \vert 0 \rangle = \int \mathcal{D}\phi\, \phi(\vec{x})\, \phi(\vec{x}')\, \Psi_0^*[\phi, t]\, \Psi_0[\phi, 0].
\end{equation}

\section{Nambu's approach in QFT}\label{sec-nambu-QFT}

Now that both Nambu's approach and Schrödinger representation in quantum field theory are properly summarized, we proceed to generalize Nambu's approach in QM to QFT.

In the light of expressions \eqref{Seq1} and \eqref{Seq2}, such a generalization is formally straightforward:
\begin{equation}
\hat{H}' \Psi[\phi, t,\tau]= i \frac{\partial}{\partial \tau} \Psi[\phi, t, \tau] \label{eq:nambu_qft}
\end{equation}
with
\begin{equation}
\hat{H}'=\hat{H}-i\frac{\partial}{\partial t}.
\end{equation}
From here on, we take $\hbar=1$,  except when an explicit reintroduction of the Planck constant is necessary.

\paragraph{Eigenvalue analysis.}

Assuming the solution of eq.\eqref{eq:nambu_qft}  of the form:
\begin{equation}
\Psi[\phi, t,\tau]= \Psi_{\lambda}[\phi, t] e^{-i \lambda \tau},
\end{equation}
we obtain a $\tau$-independent equation or an eigenvalue problem:
\begin{equation}
\hat{H}' \Psi_{\lambda}[\phi, t] = \lambda \Psi_{\lambda}[\phi, t].
\end{equation}
Notice that the operator $(-i \partial_t)$ is not  hermitian and self-adjoint, since given two states $\Psi_{1,2}$, using Eq.~\eqref{eq:inner-product-Nambu-FQFT} in appendix \ref{appedix:inner_products}, and integrating by parts, one has
\begin{equation}
\langle \Psi_1 \vert -i \partial_t \Psi_2\rangle = \int \mathcal{D}\phi  \Psi_1^{\dagger}[\phi,t]\Psi_2[\phi,t]\Big|^{+\infty}_{-\infty}+ \langle -i \partial_t\Psi_1   \vert  \Psi_2\rangle.
\end{equation}
The boundary term breaks self-adjointness. This is recovered 
by imposing the term\\
\noindent $\int \mathcal{D}\phi \Psi_1^{\dagger}[\phi,t]\Psi_2[\phi,t]$ equal at $t=\pm\infty $ -- see also \cite{Heiss_2004} for non-hermitian operators, and \cite{Bender_1998,Bender_2007} within PT-symmetric quantum mechanics .

Let us clarify how the extended proper-time dynamics translates into a shifted energy eigenvalue for the functional Hamiltonian. Starting from the generalized Nambu-like equation in the $\lambda$-sector, we have:
\begin{equation}
    (\hat{H} - i\partial_t) \Psi_{\lambda}[\phi, t] = \lambda \Psi_{\lambda}[\phi, t]\,,
\end{equation}
which can be rearranged as $\hat{H} \Psi_{\lambda}[\phi, t] = i\partial_t \Psi_{\lambda}[\phi, t] + \lambda \Psi_{\lambda}[\phi, t]$. We can map this into the standard functional Schrödinger equation by performing a phase redefinition. Let us define a new functional $\bar{\Psi}_{\lambda}[\phi, t]$ such that
\begin{equation}
    \Psi_{\lambda}[\phi, t] = e^{i \lambda t} \bar{\Psi}_{\lambda}[\phi, t]\,.
\end{equation}
Substituting this into the equation above, the $\lambda$ terms cancel out, leaving exactly the standard functional Schrödinger equation for the redefined state:
\begin{equation}
    \hat{H} \bar{\Psi}_{\lambda}[\phi, t] = i\partial_t \bar{\Psi}_{\lambda}[\phi, t]\,.
\end{equation}
If we now consider stationary states for the standard equation, we have $\bar{\Psi}_{\lambda}[\phi, t] = \Psi_{\lambda}[\phi] e^{-i \bar{E} t}$, where $\bar{E}$ is the eigenvalue of $\hat{H}$. Consequently, the original wave functional takes the form
\begin{equation}
    \Psi_{\lambda}[\phi, t] = \Psi_{\lambda}[\phi] e^{-i (\bar{E} - \lambda) t}\,.
\end{equation}
By identifying the physical, observed energy as $E \equiv \bar{E} - \lambda$ (so that $\bar{E} = E + \lambda$), we write $\Psi_{\lambda}[\phi, t] = \Psi_{\lambda}[\phi] e^{-i E t}$, which directly yields the shifted eigenvalue equation for the functional Hamiltonian:
\begin{equation}
    \hat{H}\Psi_{\lambda}[\phi] = (E+\lambda)\Psi_{\lambda}[\phi]\,.
\end{equation}
At this stage, $\Psi_{\lambda}[\phi]$ is a general wave functional describing an arbitrary field configuration. To address the particle spectrum and make contact with the standard momentum-space dispersion relations, we evaluate this equation for a single-particle eigenfunctional with  momentum $\vec{p}$. In the functional Schrödinger representation, such a state is explicitly given by~\footnote{See Eq. 10.30 in Ref.~\cite{Hatfield:1992rz}.}
\begin{equation}
\Psi_{\lambda, \vec{p}}[\phi] \equiv \tilde{\phi}(\vec{p}) \Psi_{0}[\phi] = \mathcal{N} \left( \int d^3x \, e^{-i\vec{p}\cdot\vec{x}} \, \phi(\vec{x}) \right) \Psi_{0}[\phi] \,,
\end{equation}
where $\Psi_{0}[\phi]$ is the vacuum wave functional and $\mathcal{N}$ is a normalization constant, and it is of the form~\cite{Hatfield:1992rz} 
\begin{equation}
    \Psi_{0}[\phi] = \eta \, e^{-\int\, d^3x \,  d^3 y \, \phi(\vec{x}) \, g(\vec{x},\vec{y})\, \phi(\vec{y}) }
\end{equation}
where $\eta$ is a normalization constant whose specific form is not relevant for our discussion  and 
\begin{equation}
   g(\vec{x}, \vec{y})=\int \frac{d^3 k}{(2 \pi)^3} \tilde{g}(\vec{k}) e^{i \vec{k} \cdot(\vec{x}-\vec{y})}
\end{equation}
with 
\begin{equation}
   \tilde{g}^2(\vec{k})=\frac{1}{4}\left(\vec{k}^2+m^2\right)\,.
\end{equation}
For the specific state $\Psi_{\lambda, \vec{p}}[\phi]$, the free functional Hamiltonian in \eqref{Seq2} yields the eigenvalue
$$
\omega_{\vec{p}} = \sqrt{\vec{p}^2+m^2}, 
$$
so that 
\begin{equation}
\hat{H}\Psi_{\lambda, \vec{p}}[\phi] = \omega_{\vec{p}}\,\Psi_{\lambda, \vec{p}}[\phi].
\end{equation}
The eigenvalue equation then becomes
\begin{equation}
\omega_{\vec{p}}\,\Psi_{\lambda, \vec{p}}[\phi] = (E+\lambda)\Psi_{\lambda, \vec{p}}[\phi]\,.
\end{equation}
Applying $\hat{H}$ one more time to the equation above (i.e., acting with the squared energy operator), one gets
\begin{align}
(\vec{p}^2+m^2)\Psi_{\lambda, \vec{p}}[\phi] &= (E+\lambda)^2\Psi_{\lambda, \vec{p}}[\phi] \nonumber \\
\omega_{\vec{p}}^2\Psi_{\lambda, \vec{p}}[\phi] &= (E^2+\lambda^2+2E\lambda)\Psi_{\lambda, \vec{p}}[\phi] \nonumber \\
(\lambda^2+2E\lambda+E^2-\omega_{\vec{p}}^2)\Psi_{\lambda, \vec{p}}[\phi] &= 0\,.
\end{align}
When $E^2=\omega_{\vec{p}}^2$ (on-shell states), the eigenvalue equation has two solutions $\lambda =-2\omega_{\vec{p}}$ and $\lambda=0$. The solution $\lambda=0$ corresponds to states satisfying the standard Schrödinger equation, and the negative eigenvalue we regard as on-shell states not satisfying the Schrödinger equation. 

When $E^2\neq\omega_{\vec{p}}^2$ (off-shell states), the eigenvalue equation has the two solutions
\begin{align}
\lambda_p &= E+\omega_{\vec{p}} \label{eq:lp}\\
\lambda_a &= E-\omega_{\vec{p}}\,. \label{eq:la} 
\end{align}
We shall see in the next section that these two last solutions for $\lambda$ are the ones leading to the Feynman propagator.

\paragraph{Constraint via $\tau$-Integration.}

Integrating over $\tau$ the wave functional $\Psi[\phi,t,\tau]$ produces a wave functional satisfying the standard Schrödinger equation, namely, using \eqref{Re_delta},
\begin{equation}
\Psi^R[\phi, t] \propto \text{Re}  \left(\int_{0}^{\infty} d\tau\, \Psi[\phi, t,\tau]\right)\ \propto \delta(\lambda)\, \Psi_{\lambda}[\phi, t].
\end{equation}
(The constant to transform the proportionality into an equality is inessential at this level.)

Employing the same analysis of Eq.~\eqref{eq:Schrödinger_real}, it is straightforward to show that the physical states satisfy:
\begin{equation}\label{constraint}
\hat{H}' \Psi^R[\phi, t] = 0  \hspace{2em} \Leftrightarrow \hspace{2em}   \lambda=0,
\end{equation}

\subsection{Two-Point Function and Propagator in the Nambu-like formalism}

In this formalism, in addition to the conventional time dependence of physical states, there exists a $\tau$ dependence. The projection of the state vector into the $|\phi\rangle$ basis is  as follows, 
\begin{align}
& \langle\phi \vert \Psi,t,\tau\rangle := \Psi[\phi,t,\tau] \nonumber \\
& \langle \phi \vert 0, t,\tau \rangle = \Psi_0[\phi, t,\tau],
\end{align}
such that
\begin{equation}
\langle 0,t'\equiv t,\tau'\equiv \tau \vert \hat{\phi}(\vec{x}') \hat{\phi}(\vec{x}) \vert 0,t\equiv 0,\tau\equiv 0 \rangle = \int \mathcal{D}\phi\, \phi(\vec{x})\, \phi(\vec{x}')\, \Psi_0^*[\phi, t, \tau]\, \Psi_0[\phi, 0, 0],
\end{equation}
where in the R.H.S. we have inserted the completeness relation $\int\mathcal{D}\phi |\phi\rangle\langle\phi| = \mathbf{1}. 
$

Assuming the usual eigenvalue decomposition:
\begin{equation}
\Psi_0[\phi, t, \tau] = \Psi_0[\phi,t]\, e^{-i \lambda \tau},
\end{equation}
we find:
\begin{equation}
\langle 0, t, \tau \vert \hat{\phi}(x') \hat{\phi}(x) \vert 0, 0, 0 \rangle = e^{-i \lambda \tau} \int \mathcal{D}\phi\, \phi(\vec{x})\, \phi(\vec{x}')\, \Psi_0^*[\phi,t] \Psi_0[\phi,0].
\end{equation}
which, integrated over $\tau$ and extracting the real part to enforce the constraint as discussed for the $\lambda=0$ sector, gives
\begin{equation}
\text{Re} \left( \int_0^\infty d\tau\, \langle 0, t, \tau \vert \hat{\phi}(\vec{x}') \hat{\phi}(\vec{x}) \vert 0, 0, 0 \rangle \right) \propto \delta(\lambda) \int \mathcal{D}\phi\, \phi(\vec{x})\, \phi(\vec{x}')\, \Psi_0^*[\phi,t] \Psi_0[\phi,0].
\end{equation}
Now recall that isolating $\delta(\lambda)$ -- i.e., enforcing $\lambda=0$ -- selects the standard Schrödinger functional equation, thus leading to the standard propagator. Then, going to energy-momentum space, one can construct the proper-time transition amplitude. To correctly reproduce the time-ordered causal structure of QFT, one must symmetrically include both the particle and antiparticle off-shell branches. This is achieved by introducing a mixing parameter $\rho \in [0,1]$, akin to a Feynman parameter, which linearly interpolates between the two solutions. The amplitude is then written as:
\begin{align}
& \langle 0, t, \tau \vert \hat{\phi}(\vec{x}') \hat{\phi}(\vec{x}) \vert 0, 0, 0 \rangle \rightarrow
\langle 0, p_0, \tau \vert \hat{\phi}(\vec{p}) \hat{\phi}(0) \vert 0, 0, 0 \rangle \nonumber \\
& = -i \, \tau \,\int_0^1 d\rho\,e^{-i\rho\lambda_p\tau -i (1-\rho)\lambda_a\tau},
\end{align}
with $\lambda_p = p_0-\omega_{\vec{p}}$ and $\lambda_a = p_0+\omega_{\vec{p}}$, which are precisely the expressions found in \eqref{eq:lp} and \eqref{eq:la} with $p_0 =E$. The parameter $\rho$ essentially ensures that both positive and negative frequency modes are on the same footing. Integrating over $\tau$ one recovers exactly the standard Feynman propagator
\begin{equation}\label{rep1}
-i\int_0^{\infty}d\tau \, \tau \,\int_0^1 d\rho\,e^{-i\rho\lambda_p\tau -i (1-\rho)\lambda_a\tau}=\frac{i}{p^2-m^2+i\epsilon}.
\end{equation}
(Note: we implicitly assume the standard $i\epsilon$ prescription to ensure the convergence of the $\tau$ integration at infinity, as detailed in Appendix \ref{appendix:newderivation}).

On the other hand, Schwinger's representation is
\begin{equation}\label{rep2}
 \langle 0 \vert T \hat{\phi}(p) \hat{\phi}(0) \vert 0 \rangle= \frac{i}{p^2-m^2} 
= \int_0^\infty ds\, e^{i s (p^2 - m^2 +i 0^+)}.
\end{equation}
The expressions \eqref{rep1} and \eqref{rep2} are equivalent integral representations of the Feynman propagator. In Schwinger’s approach, the integral involves the proper time squared, whose generator is the four-momentum $p^2$, and we work with the proper time, whose generators are the eigenvalues of $\lambda$. In contrast, the integration limits of \eqref{rep1} and \eqref{rep2} are in correspondence, which will be relevant later. We elaborate on the interpretation of the proper-time representation of the Feynman propagator in Appendix \ref{appendix:interpretation}.

We have analyzed the free model in full detail, which serves as the basis for quantization aiming to treat the interaction as a perturbation. The canonical quantization procedure is performed on the free Lagrangian, whose exact solvability allows for a well-defined Fock space construction. Interactions are subsequently introduced as perturbations built upon this free, quantized structure. This methodology follows the conventional formulation of canonical quantization in QFT. Once an interaction is turned on,
\begin{equation}\label{Hint}
H_{\text{int}} = \int d^3 x \, \frac{\alpha}{4!} \, \phi(\vec{x})^4,
\end{equation}
the coupling $\alpha$ evolves according to renormalization group equations (RGEs). Usually, renormalization involves introducing a reference energy scale $\mu$ to absorb the cutoff dependence in loop integrals, and the RGEs describe how the coupling $\alpha(\mu)$ evolves with $\mu$ -- see Appendix \ref{appendix:RGE}.

\section{Introducing a physical $\tau_{\min} > 0$: Constraint violation}\label{sec-constraint}

This central section aims to introduce a minimal proper time, $\tau_{min}>0$. Notice that the nonzero $\tau$ used in the previous subsection was only a regulator, thus to be fixed to zero after any calculations. On the contrary, here we promote it to a physical scale,  $\tau_{min}$.

Identifying $\tau_{min}$ with a fundamental physical scale is natural, being proper time intrinsically Lorentz invariant.
Heuristically, if $\tau$ denotes the proper time along a timelike worldline ($c=1$),
\begin{equation}
\tau^2 = \Delta t^2 - \Delta x^2 \ge \tau_{\min}^2 = s_{min},
\end{equation}
then $\tau_{\min}$ immediately implies that no physical process can probe arbitrarily
short intervals in spacetime. In particular, if $\Delta x=0$ one
finds $\Delta t \ge \tau_{\min}$. 
Thus, a minimal proper time entails a
minimal temporal resolution. Naturally, we identify $\tau_{min}$ with the Planck's time $\sqrt{\frac{\hbar\,G}{c^5}}$.

\subsection{Relaxing the constraint and exotic states}

Naively, the introduction of a minimal proper-time $\tau_{min}$ makes the right-hand side of Eq. \eqref{eq:nambu_qft} ill-defined. To solve this issue, one can resort to an analogy with the approach of   \cite{Maiezza:2025bzg}. The functional $\Psi[\phi,t,\tau]$ can be made differentiable by formally introducing a $\tau_{min}$ dependence in the functional  $\Psi[\phi,t,\tau]\to \Psi[\phi,t,\tau\,|\,\tau_{min}]$ via  a smoothing function. In other words, we regularize $\Psi$:
\begin{equation}\label{reg_smooth}
\Psi_{reg}[\phi,t,\tau] = \int_0^\infty d\tau' S(\tau-\tau'\vert \tau_{min}) \Psi[\phi,t,\tau'] \,,
\end{equation}
where the subscript '$reg$' denotes the regularized functional, and $S$ is a smoothing kernel. 
For the sake of notation, in what follows we will omit the subscript '$reg$' and the explicit $\tau_{min}$ dependence. Changes in the scale $\tau_{min}$ are, as usual, governed by the renormalization group equation, as discussed in appendix \ref{appendix:RGE}. 

Once the functional $\Psi[\phi,t,\tau]$ is properly regularized, one can consider the generalized functional Schrödinger equation with a parameter $\lambda$, 
\begin{equation}
\left(\hat{H} - i \frac{\partial}{\partial t}\right) \Psi_\lambda[\phi,t] = \lambda \Psi_\lambda[\phi,t].
\label{eq:general_shift}
\end{equation}
Rearranging, we have
\begin{equation}
\hat{H} \Psi_\lambda[\phi,t] = i \frac{\partial}{\partial t} \Psi_\lambda[\phi,t] + \lambda \Psi_\lambda[\phi,t].
\end{equation}
Defining a new wavefunction
\begin{equation}\label{eq:phase_redef}
\tilde{\Psi}_{\lambda}[\phi,t] = e^{-i \lambda t} \Psi_\lambda[\phi,t],
\end{equation}
we find that $\tilde{\Psi}$ satisfies the standard Schrödinger equation
\begin{equation}\label{eq:standard_Schrödinger}
\hat{H} \tilde{\Psi}_{\lambda}[\phi,t] = i \frac{\partial}{\partial t} \tilde{\Psi}_{\lambda}[\phi,t].
\end{equation}
Since both $\tilde{\Psi}_{\lambda}[\phi,t]$ and $\tilde{\Psi}_{\lambda=0}[\phi,t]$ satisfy Schrödinger equation it must be that
\begin{equation}\label{eq:Alambda}
    \tilde{\Psi}_{\lambda}[\phi,t] = A_{\lambda} \, \Psi_{\lambda=0}[\phi,t] .
\end{equation}
In the standard formulation, the integration over the auxiliary proper-time parameter $\tau$ enforces the constraint on physical states via the Dirac delta:
\begin{equation}
\int_0^\infty d\tau\, e^{-i\lambda\tau} = \pi\delta(\lambda) - i\, \mathcal{P}\left(\frac{1}{\lambda}\right),
\end{equation}
being  $\mathcal{P}$ the principal value, and where the real part selects the physical states satisfying the constraint:
\begin{equation}
\text{Re}\left[\int_0^\infty d\tau\, e^{-i\lambda\tau}\right] = \pi\delta(\lambda).
\end{equation}

To promote the introduction of a minimal proper time $\tau_{\min}$ from a mere regularization device to a structural assumption of the formalism, we postulate that the proper-time representation of the theory is not operationally defined below a fundamental lower bound $\tau_{\min} > 0$. In other words, phenomena associated with scales shorter than $\tau_{\min}$ are assumed to lie outside the domain of validity of the effective description.
	
Under this assumption, the strict projection onto the physical sector obtained by integrating down to $\tau = 0$ — which yields an exact $\delta(\lambda)$ constraint — is interpreted as an idealized limit. At finite resolution, the lower bound $\tau_{\min}$ prevents this exact projection, replacing it with a smeared constraint in which modes with $\lambda \neq 0$ are not completely eliminated.
	
Accordingly, the proper-time integral defining the physical sector must explicitly reflect this minimal scale. We therefore decompose
\begin{equation}
\int_0^\infty e^{-i\lambda\tau}\, d\tau
=
\int_0^{\tau_{\min}} e^{-i\lambda\tau}\, d\tau
+
\int_{\tau_{\min}}^\infty e^{-i\lambda\tau}\, d\tau \,,
\end{equation}
and interpret the contribution from the interval $[0,\tau_{\min})$ as lying beyond the regime of validity of the effective theory.

Since the full integral yields the delta function, we can write:
\begin{equation}
\text{Re}\left[\int_{\tau_{\min}}^\infty e^{-i\lambda\tau} d\tau\right] = \pi\delta(\lambda) - \text{Re}\left[\int_0^{\tau_{\min}} e^{-i\lambda\tau} d\tau\right].
\end{equation}
To leading order in $\tau_{\min}$, we expand the second term:
\begin{align}
&\int_0^{\tau_{\min}} e^{-i\lambda\tau} d\tau = \frac{1 - e^{-i\lambda\tau_{\min}}}{i\lambda} \quad \Rightarrow \quad \nonumber \\
&\text{Re}\left[\cdot\right] = \frac{\sin(\lambda \tau_{min})}{\lambda} = \tau_{\min} + \mathcal{O}(\tau_{\min}^3),
\end{align}
valid for $\lambda \tau_{min}<<1$ (the validity of the expansion will be further refined for a specific form of $f(\lambda)$).

Therefore, the constraint is relaxed in a controlled way:
\begin{equation}\label{approx_viol}
\text{Re}\left[\int_{\tau_{\min}}^\infty e^{-i\lambda\tau} d\tau\right] = \pi\delta(\lambda) - \tau_{\min} + \mathcal{O}(\tau_{\min}^3),
\end{equation}
having used the linear expansion in $\tau_{min}$ above.

This implies that the physical state is no longer strictly constrained to $\lambda = 0$, but instead admits a small support around it. Exotic states with $\lambda \neq 0$ are allowed, though suppressed. The expansion in $\tau_{min}$ is justified by its necessary smallness (since it is related to the inverse Planck mass). In the next subsection, we consider this point.

\subsection{Superposition of Eigenstates}

Once non-zero values for $\lambda$  are allowed, the wave functional is the general superposition: 
\begin{align}\label{superposition}
& \Psi[\phi,t,\tau]=\int_{-\infty}^\infty d\lambda\, f(\lambda)\, \Psi_\lambda[\phi,t]\, e^{-i \lambda \tau}   \nonumber \\
&\Psi_{R}[\phi,t] = \int_{\tau_{min}}^\infty d\tau \int_{-\infty}^\infty d\lambda\, f(\lambda)\, \Psi_\lambda[\phi,t]\, e^{-i \lambda \tau}
\end{align}
where the subscript $R$ denotes 'Real-world', such that when $\tau_{min}\rightarrow 0$ one has the original constrained $\Psi_{(\lambda=0)}$. Similarly, in opposite logical direction, when one has simply $f(\lambda)=\delta(\lambda)$ one recovers  $\Psi^R=\Psi_{(\lambda=0)}$ and $\tau_{min}=0$. Notice that this $\tau_{min}$ dependence in $ f(\lambda)$ emanates from the regularization of the functional, namely, \eqref{reg_smooth}. 

The expression \eqref{superposition} resembles the standard QM expression,
\begin{equation}
\psi(\vec{x},t) = \sum_i c_i\, \psi_i(\vec{x})\, e^{-i E_i t}, \hspace{2em}  c_i= \int dx^3 \psi^*_i(\vec{x}) \, \psi(\vec{x},0).
\end{equation}
In full analogy with QM and the above coefficients $c_i$, the expression for $f(\lambda)$ reads
\begin{equation}
f(\lambda) = \int \mathcal{D}\phi \, \Psi[\phi,0,0] \, \Psi_\lambda^*[\phi,0].
\end{equation}
Using \eqref{eq:Alambda} and \eqref{approx_viol}, the real part of the expression in \eqref{superposition}  at  leading order in $\tau_{min}$ takes the form 
\begin{equation}\label{gen1}
\Psi_{R}[\phi,t] = \Psi_{\lambda=0}[\phi,t] - \tau_{min} \Psi_{\lambda=0}[\phi,t] \int_{-\infty}^{\infty} d\lambda f(\lambda) e^{i\lambda t}  + \mathcal{O}(\tau_{min}^2)\,.
\end{equation}
where we have absorbed the constants into the redefinitions: 
\begin{align}
    \pi f(0) A_{\lambda=0}\Psi_{\lambda=0}[\phi,t] &\rightarrow \Psi_{\lambda=0}[\phi,t]\, \nonumber \\
     \frac{A_{\lambda}}{\pi f(0) A_{\lambda=0}}f(\lambda) &\rightarrow f(\lambda) .
\end{align}
Finally, we arrive at
\begin{equation}\label{gen3}
\Psi_{R}[\phi,t] = \Psi_{\lambda=0}[\phi,t] \left[ 1- \tau_{min} \tilde{f}(t) \right]+ \mathcal{O}(\tau_{min}^2),
\end{equation}
where
\begin{equation}
\tilde{f}(t) := \int_{-\infty}^{\infty} d\lambda f(\lambda) e^{i \lambda t}. 
\end{equation}
The relevant outcome is that \eqref{gen3} implies a violation of unitarity in the theory.

\section{Features and Implications of the Minimal Proper Time}\label{sec-emergent}

In this section, we discuss some key points and implications.

\subsection{On the function $f(\lambda)$}

The function $f(\lambda)$ plays a central role in the proper-time Schrödinger framework, as it encodes the spectral weight of states with non-zero eigenvalue $\lambda$.  However, it is important to emphasize that $f(\lambda)$ is not calculable from first principles within the present formalism. This is not a flaw, but a structural feature: just as in quantum mechanics, one computes the eigenstates of a Hamiltonian, but the actual physical state must be externally prepared and decomposed in terms of those eigenstates. Here $f(\lambda)$ reflects the preparation of the initial wavefunctional.
\vspace{0.3cm}
\noindent
The only requirement on $f(\lambda)$ is that it be consistent with the standard limit. Specifically, when the minimal proper time $\tau_{\min} \to 0$, the theory must reduce to the conventional Schrödinger representation, which imposes the constraint $\hat{H}'\Psi = 0$. This corresponds to selecting $\lambda = 0$, and therefore:
\begin{equation}
f(\lambda) \to \delta(\lambda) \quad \text{as} \quad \tau_{\min} \to 0.
\end{equation}
A natural and physically motivated choice for $f(\lambda)$ is a properly normalized Gaussian:
\begin{equation}
f(\lambda) = \frac{1}{\sqrt{2\pi\sigma^2}}\, e^{-\frac{\lambda^2}{2\sigma^2}},
\end{equation}
where $\sigma$ is a small parameter controlling the spread around $\lambda = 0$. As $\sigma$ approaches zero, this function approaches the Dirac delta function:
\begin{equation}
\lim_{\sigma \to 0} f(\lambda) = \delta(\lambda),
\end{equation}
thus recovering the constrained theory. This choice also allows for analytic control over the time-dependent function:
\begin{equation}\label{ft}
\tilde{f}(t) := \int_{-\infty}^{\infty} d\lambda\, f(\lambda)\, e^{i\lambda t} = e^{-\frac{1}{2}\sigma^2 t^2},
\end{equation}
which is real (energy is conserved). 
\vspace{0.3cm}
\noindent

In summary, $f(\lambda)$ is an input to the theory, its only constraint is consistency with the standard limit, and the Gaussian form provides a physically reasonable and mathematically tractable choice.

\subsection{Unitarity violation and deformation of CCR}

The unitarity violation in \eqref{gen3} implies that the evolution operator can be written as,
\begin{equation}\label{Uviolationgen}
\tilde{U}(t) = U \left(1 - \epsilon \tilde{f}(t)\right) + \mathcal{O}(\epsilon^2),
\end{equation}
where $U$ is the standard unitary, time-evolution operator, while the remaining part describes unitarity violation during temporal evolution in terms of the dimensionless parameter:
\begin{equation}
\epsilon = \sigma \tau_{min},
\end{equation}
being $\sigma$ defined in \eqref{ft}.

The expressions in \eqref{Uviolationgen} and \eqref{gen3} have to be further clarified in terms of the temporal generator. 

So let us rewrite \eqref{gen3} as,
\begin{align}
\Psi_R (t) =& U(t,t_0) \Psi_R (t_0) \left(1 - \epsilon \tilde{f}(t)\right) + \mathcal{O}(\epsilon^2) \nonumber \\
&:= U(t,t_0) \Psi_R (t_0)\, g(t),
\end{align}
where -- we recall,
\begin{equation}
U(t,t_0)= \mathcal{T} \exp\left(\int_{t_0}^{t} H(t') dt'   \right).
\end{equation} 
Now defining,
\begin{equation}
S(t):= -\log g(t)  \hspace{1em} \text{or} \hspace{1em} g(t)= e^{-S(t)},
\end{equation} 
one has,
\begin{equation}
\tilde{U}(t,t_0)= e^{-\left[S(t)-S(t_0)\right]} U(t,t_0),
\end{equation} 
which, upon differentiation in time, leads to the equation:
\begin{equation}
i \partial_t \tilde{U}(t,t_0)= \left(H-i \dot{S}(t)\right) \tilde{U}(t,t_0),
\end{equation} 
where dot denotes time derivative. Next, defining $K(t):= \dot{S}(t)$, we rewrite it as
\begin{equation}
i \partial_t \tilde{U}(t,t_0)= \left(H-i K(t)\right) \tilde{U}(t,t_0).
\end{equation} 
resulting in an effective non-self-adjoint Hamiltonian,
\begin{equation}
H_{eff}= H-i K.
\end{equation} 
Identifying it as the effective generator guarantees
correct time compositions for $\tilde{U}$. Additionally, this connects naturally to open quantum systems since the dynamic effectively resembles a Lindblad-type evolution with controlled decoherence. We will not build an explicit Lindblad dynamic in this work.

In Heisenberg representation of (standard, with no minimal proper time) QFT, one has then,
\begin{equation}\label{standard-trans}
\hat{\phi}(\vec{x},t) = U(t)^\dag \hat{\phi}(\vec{x},0) U(t), \quad \hat{\pi} = U(t)^\dag \hat{\pi}(\vec{x},0) U(t),
\end{equation}
which preserves equal-time canonical commutation relations (CCRs):
\begin{equation}\label{CCR}
[\hat{\phi}(\vec{x}), \hat{\pi}(\vec{y})]  = i\hbar\, \delta(\vec{x} - \vec{y}).
\end{equation}
Conversely, in the case of $\tau_{min}>0$, one has to modify
\eqref{standard-trans} as
\begin{equation}
U\mapsto \tilde{U},
\end{equation}
and CCRs are no longer preserved due to the non-unitarity of the transformations. Specifically, one obtains,
\begin{equation}\label{deforme}
[\hat{\phi}(\vec{x}), \hat{\pi}(\vec{y})] = i\hbar\, \delta(\vec{x} - \vec{y}) \left(1 - 4 \epsilon  \tilde{f}(t)\right) + \mathcal{O}(\epsilon ^2),
\end{equation}
where the factor 4 comes from the application of $\tilde{U}^\dag$ and  $\tilde{U}$ two times and keeping the linear terms in $\tilde{f}(t)$.

\paragraph{Implication of the modified canonical commutation relation.}
The residual non-unitarity in \eqref{Uviolationgen} implies that the representations of the free and interacting field algebras are generically nonequivalent. As a specific application of \eqref{deforme}, one has in the interaction picture,
\begin{equation}
[\hat{\phi}_I(\vec{x}), \hat{\pi}_I(\vec{y})] = [\hat{\phi}_F(\vec{x}), \hat{\pi}_F(\vec{y})] \left(1 - 4 \epsilon  \tilde{f}(t)\right) + \mathcal{O}(\epsilon ^2),
\end{equation}
where the subscripts I and F denote interacting and free field, respectively.

This structural feature of the proper-time formalism is naturally in agreement with  Haag's theorem \cite{Haag:1955ev}. Haag's theorem establishes that if (1) the theory is translationally invariant and (2) the free and interacting field representations are related by a unitary transformation, then the $n$-point correlation functions of the free and interacting theories must be identical—in direct contradiction with the results of renormalized perturbation theory. In the present framework, the evolution operator is intrinsically non-unitary due to the minimal proper time, thereby violating assumption (2) \footnote{See \cite{Segreto:2025cbv} for an analysis on the unitary equivalence within the generalized uncertainty principle theories.}. Consequently, Haag's theorem does not apply, and the free and interacting correlation functions can legitimately differ, as required by perturbative calculations. This should not be interpreted as a resolution of the foundational issues underlying Haag's theorem in standard perturbation theory, but rather as a framework where the standard assumptions are modified, placing it outside the theorem's domain of applicability. What is far from trivial in the  present framework is that it is able to accommodate all standard QFT results at energies much lower than the scale $1/\tau_{min}$ -- since renormalizability is preserved --  while the  Lorentz and unitarity  violation is always suppressed by powers of  $\tau_{min}$.

\subsection{Effective Planck constant}

The deformation in \eqref{deforme} can be interpreted as an effective, time-dependent Planck constant:
\begin{equation}\label{hrun}
\hbar_{eff}(t) := \hbar \left(1 - 4 \epsilon  \tilde{f}(t)\right) + \mathcal{O}(\epsilon ^2).
\end{equation}
It  vanishes smoothly in the limit $\epsilon \to 0$ ($\tau_{\min}\to 0$). Therefore, the underlying algebraic structure of the theory remains intact, with only controlled deviations induced by the minimal proper time. Here, we aim to provide a possible interpretation. 

Due to the form of $\tilde{f}(t)$ in \eqref{ft}, the minimal value of $\hbar_{\text{eff}}$ in \eqref{hrun} occurs at $t = 0$. Moreover, this deviation from the constant $\hbar$ is small, being suppressed by the parameter $\tau_{\min}$.

We observe that the theory contains a degree of freedom that can be exploited to clarify the meaning of $\hbar_{\text{eff}}(t)$. The key point is that the Nambu-like constraint can be imposed with a Dirac delta function having arbitrary normalization. Regardless of any prefactor in front of $\delta(\lambda)$ in \eqref{constraint}, the constraint \textit{a la} Nambu still selects $\lambda = 0$. Similarly, in the presence of a soft violation due to $\tau_{\min} > 0$, the function $f(\lambda)$ can be normalized with an arbitrary constant. Thus it reflects on $\tilde{f}(t)$ as a rescaling: $\tilde{f}(t) \rightarrow N \tilde{f}(t)$, where $N$ denotes the normalization factor. As a result, equation \eqref{hrun} becomes:
\begin{equation}\label{hrun2}
\hbar_{eff}(t) := \hbar \left(1 - 4 N \epsilon \tilde{f}(t)\right) + \mathcal{O}(\epsilon^2).
\end{equation}
If $4 N \epsilon \approx 1$, and since $\tilde{f}(0) = 1$, we find $\hbar_{\text{eff}}(0) \approx 0$. This implies that the effective Planck constant can vanish, or become significantly smaller than the measured value, at very short times $t < \tau_{\min}$. When $h_{eff}\sim 0$, quantum fluctuations are suppressed, and the system (e.g., a self-interacting real scalar field) will start behaving deterministically. Decoherence, therefore, appears at short times ($t\propto 1/\tau_{min}\sim M_{Pl}$, being the latter the Planck mass), while quantum behavior emerges at larger times (or lower energies).    

A couple of remarks and comments are now in order: 
\begin{enumerate}
\item 
The condition $4 N \epsilon \approx 1$ is one possible way to fix a free constant of the theory via a physical principle, namely, determinism at the fundamental level. In fact, it enables a reasonable motivation for a time-dependent Planck's constant, realizes a decoherence at short times,  with a smooth transition from  quantum to deterministic behavior. This is an interpretative choice and does not follow from first principles, although it remains fully compatible with the structure of the framework. 

\item 
We can also offer a heuristic 
argument in the opposite direction. If nature is fundamentally deterministic with quantum behavior only  emergent, one may expect non-unitarity at the transition scale, since in the end,  information is not destroyed, but rather dispersed into deterministic degrees of freedom that might not accept a quantum description. Further analyses, beyond the scope of this work, may be dedicated to investigating these aspects. 
\end{enumerate}

The latter point and the expression of $\hbar_{eff}$ in \eqref{hrun2} might provide a natural connection to scenarios in which nature is assumed to be fundamentally deterministic at energies larger or of the order of the Planck's mass high, whereas quantum mechanics arises as an emergent phenomenon at lower energies -- see, for example, the cellular automaton proposed by 't Hooft in \cite{hooft2015cellularautomatoninterpretationquantum}. Here, the Hilbert-space structure and probabilistic interpretation arise as effective constraints on coarse-grained deterministic states. In this framework, unitarity is not a fundamental property but an emergent one: at the microscopic level, the dynamics is deterministic, while an effective unitary evolution appears only after representing the deterministic map as a permutation operator acting on a Hilbert-space basis. Different perspectives and recent developements on deterministic models are discussed in \cite{Adlam:2023fmq,Arroyo:2024saq,Palmer:2025guq} \footnote{In a radically different context, the authors \cite{Dvali:2010jz,Dvali:2011aa} proposed the concept of 'classicalization', where high-energy scattering amplitudes are dominated by extended classical field configurations rather than by new weakly-coupled quantum degrees of freedom, providing a classical UV completion that preserves unitarity.}.  

Our framework differs in essential aspects. The deterministic proposals can be seen as a top-bottom, in energy scale, approach; our setting is a bottom-top approach since we start from a quantum description, such that QFT must be understood as an effective theory. The minimal proper-time deformation we introduce preserves renormalizability while inducing a soft relaxation of unitarity, parametrized by $\epsilon$ in Eq.~\eqref{Uviolationgen}.  The resulting deviations are suppressed by powers of the heavy scale, ensuring that for $E \ll M$ the effective $S$-matrix remains unitary and Lorentz invariant to any experimental precision.  This allows quantum mechanics to emerge dynamically as a low-energy constraint, while $\hbar_{\text{eff}}\!\to\!0$ at high energies leads to a deterministic UV regime.  In this sense, our construction realizes a controlled and renormalizable route toward a deterministic ultraviolet completion, distinct from both the cellular–automaton approach and the classicalization scenario, yet conceptually aligned with the idea that unitarity and quantumness may be effective low-energy properties rather than fundamental principles.

\subsection{Effect on the running coupling behavior at high energies}

The presence of a minimal proper time $\tau_{\min}$ modifies the propagator by introducing an exponential suppression of high-energy modes. After Wick rotation to Euclidean space, the propagator reads:
\begin{equation}\label{prop_dumped}
G_E(p^2;s_{\min}) = \int_{s_{\min}}^\infty ds\, e^{-s(p_E^2 + m^2)} = \frac{e^{-s_{\min}(p_E^2 + m^2)}}{p_E^2 + m^2}.
\end{equation}
At large $p_E^2$, the exponential factor $e^{-s_{\min} p_E^2}$ strongly suppresses UV contributions, effectively taming divergences and improving the ultraviolet behavior of the theory. Remaining in $\phi^4$-model, this mechanism ensures that loop integrals become finite and suggests the absence of a Landau pole, as in \cite{Maiezza:2022xqv}. The reason is that the running of the coupling is no longer log-dominated at high energy, as in the standard case. Specifically, it is easy to work out the behavior of the four-point function $\alpha/4!\, \phi(x)^4$ model, in the relevant regimes, using the propagator in \eqref{prop_dumped}:
\begin{align}
&\alpha(\mu)\approx \alpha(\mu_0) \left(1+ \alpha(\mu_0)\, \beta_0  \log(\frac{\mu}{\mu_0}) - \alpha(\mu_0) s_{min} \mu^2\right)   \hspace{1.5em} \mu\ll s_{min}^{-1/2}=\tau_{min}^{-1} \nonumber\\
& \alpha(\mu)\approx \alpha(s_{min}^{-1/2}) \left(1 - \beta_0\,\alpha(s_{min}^{-1/2}) \frac{e^{-2 s_{min} \mu^2}}{4 s_{min} \mu^2} \right)     \hspace{3.6em} \mu\gtrsim s_{min}^{-1/2}=\tau_{min}^{-1} , \label{alphaUV}
\end{align}
where $\beta_0=3/(16\pi^4)$. The first line reproduces the standard, log behavior, as it must be. The second one shows that the coupling rapidly reaches a plateau in the UV, rendering $\phi^4$-model asymptotically safe. It effectively becomes scale invariant at very high energy.

\begin{figure}[t]
    \centering
    \includegraphics[width=0.6\textwidth]{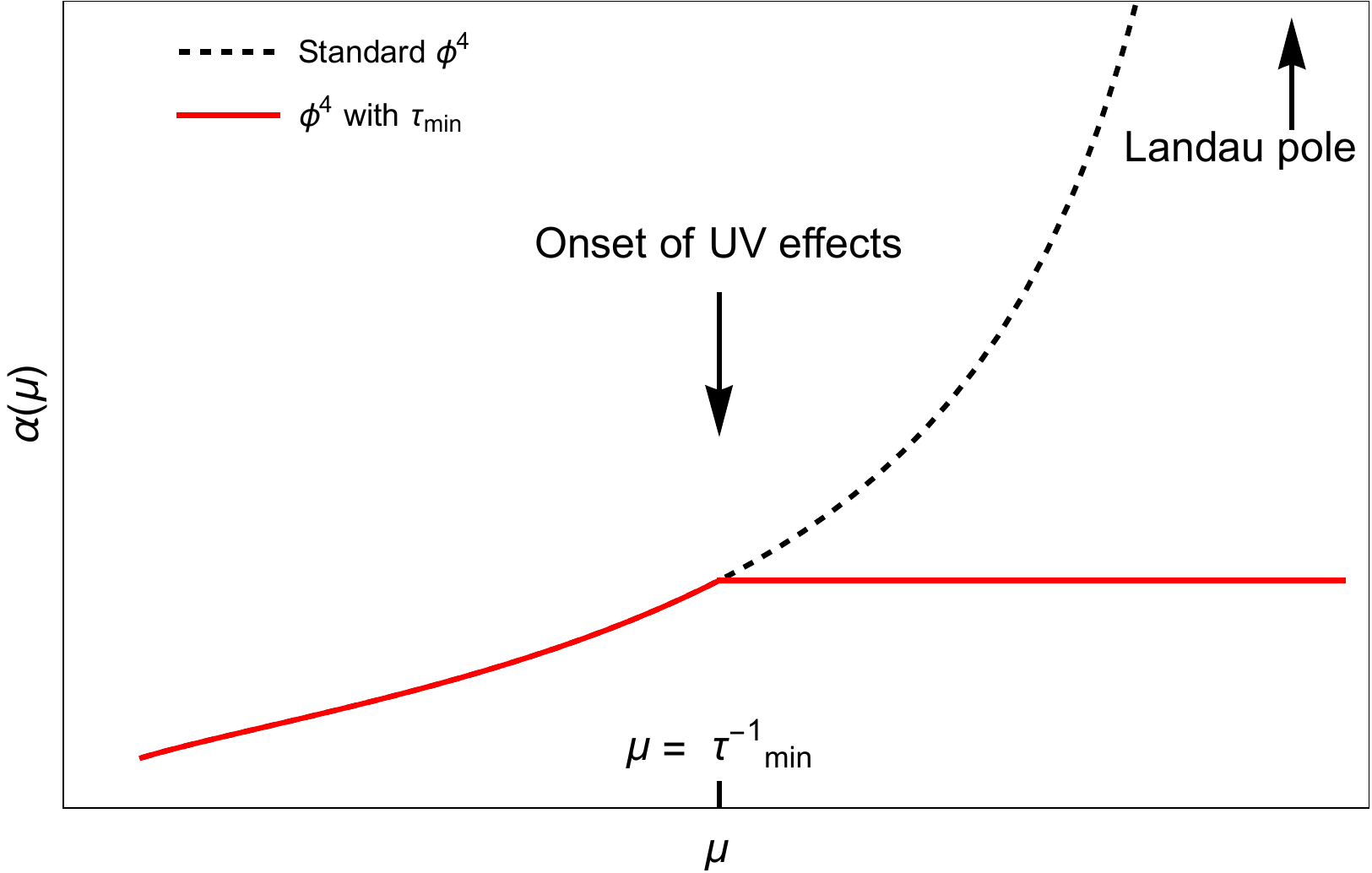}
    \caption{Qualitative comparison of the standard $\phi^4$ running coupling  and its version with a minimal proper time. 
    The two coincide in the infrared, while the new UV effects become relevant above the scale $\tau_{min}$. 
    In contrast to the standard prediction, which develops a Landau pole, 
    the running in the new framework remains finite in the ultraviolet.}
    \label{fig:uv_running}
\end{figure}

Some implications are now highlighted.

\begin{enumerate}

\item A conceptual obstacle for perturbative QFT is the presence of the UV renormalons which destroy the Borel resummability of the perturbative series~\cite{tHooft:1977xjm}. The renormalons arise from the logarithm behavior of the running coupling in deep UV -- this is particularly manifest from the resurgent approach in \cite{Bersini:2019axn,Maiezza:2023mvb} -- causing the appearance of factorial divergent contribution in the power series. In the present framework,
the modification due to $\tau_{min}$ in \eqref{alphaUV} modifies the standard logarithm behavior as also illustrated in Fig. \ref{fig:uv_running}, then eliminating the source of the UV renormalon ambiguities\cite{Maiezza:2022xqv}. This is a  nontrivial result: the absence of renormalons indicates that the perturbative expansion remains well-controlled. Combined with the automatic and guaranteed asymptotic safety of the theory, this ensures that the renormalization procedure is internally consistent and free from pathological ambiguities at arbitrarily high energies.

Additionally, the absence of logarithmic running of the coupling in the ultraviolet (UV) regime invalidates ’t Hooft’s argument for the so-called horn-shaped singularities in the Green functions~\cite{tHooft:1977xjm}. As shown in \cite{Maiezza:2022xqv}, without logarithmic behavior the perturbative expansion does not exhibit super-exponential growth in Borel space. This removes a potential source of pathological UV behavior in perturbation theory and significantly reinforces the theoretical consistency of the framework at short distances, while preserving all standard QFT predictions at low energies.

These considerations further motivate our strategy: we first analyze the free-field theory in detail and subsequently introduce the interaction term in \eqref{Hint}, to be treated perturbatively.

\item In the deep UV regime, $E > \tau_{\min}^{-1}$, loop effects should be regarded as a formal extrapolation within our framework. Indeed, the progressive vanishing of $\hbar_{\text{eff}}$ in the UV, see \eqref{hrun2}, suppresses quantum corrections and therefore loop contributions. This behavior is consistently reflected in the exponential damping of the propagator in \eqref{prop_dumped}, which provides an effective description of the same mechanism. As a result, the $\beta$-function asymptotically vanishes and exact scale invariance is recovered in the ultraviolet.

\item The regular UV behavior can be naturally interpreted as a \emph{dimensional reduction} -- see \cite{Carlip_2017} for a review. Specifically, the term $\exp\left[-s(p_E^2 + m^2)\right]$ works as an entire function similarly to \cite{Tomboulis:1997gg,Maiezza:2022xqv,Briscese_2024,Maiezza:2025bzg} -- an evolving dimension is discussed in \cite{Afshordi_2014,Dai_2014,ANCHORDOQUI_2012}, and the non-integer effective dimension can then be interpreted as fractal spacetime \cite{Modesto_2009,Benedetti_2009,Calcagni_2010,Calcagni:2010bj,CALCAGNI_2013}. The underlying idea is that the effective dimension, different from the topological one ($D=4$), may run with the energy and lower in the UV. The UV behavior of the Green function is the one to measure the effective dimensionality \cite{Maiezza:2025bzg}. Similarly, one can 
write a modified heat kernel, such that the corresponding spectral dimension is variable, vanishing at UV \cite{ Modesto_2010,Eckstein_2020}.

\end{enumerate}

\section{Conclusion}\label{sec-end}

Exploiting the Schrödinger representation in quantum field theory and extending Nambu’s approach from quantum mechanics to quantum field theory, we have explored a framework incorporating a minimal proper time $\tau_{min}$. This minimal proper time acts as a Lorentz invariant ultraviolet regulator. In this framework, the dynamics comes from the evolution in proper time, and  the quantum mechanical  time evolution is interpreted as a constraint,  rather than a fundamental dynamical evolution. 

The introduction of $\tau_{min}$ leads to the main consequences: 
\begin{enumerate}

\item  The appearance of a time-dependent Planck constant able to accommodate a vanishing value (deterministic)  at time scales smaller than $\tau_{min}$, and an emergent quantum behavior at later times. 

\item  The proper-time suppresses high-energy modes, while keeping all the standard QFT predictions at lower energies,  leading to a smooth UV transition. This can be  reinterpreted as an effective dimensional reduction at short distances. This resonates with phenomena expected in quantum gravity, where spacetime dimensionality flows from four to lower values in the deep ultraviolet and exhibits automatic asymptotic safety.

\item A mild violation of unitarity at very high energies, while recovered at large distances of the order of the quantum regime. The violation is achieved through an effective deformation of canonical commutation relations through a non-unitary Dyson operator. The latter implies that free and interacting fields are not unitarily equivalent,  in agreement with  Haag’s theorem.

\end{enumerate}

A brief, explicit comparison with the related literature may be helpful. Often, the minimal length is introduced in terms of the generalized uncertainty principle or modified dispersion relation \cite{Hossenfelder_2013}. More recently, the minimal length has been presented as a momentum-space cutoff that acts independently of the system’s dynamics \cite{Bosso_2023}. While the present proposal shares with these references the motivation, namely a scale related to the completion of gravity, it departs in several aspects. The minimal proper time is introduced in QFT to violate Nambu's  constraint -- preserving renormalizability -- in such a way that its  effects are always suppressed by inverse powers of the Planck scale. The violation of the aforementioned constraint also implies the  deformation of the canonical commutation relations, again suppressed by inverse powers of the Planck scale.

Moreover, we interpret $\tau_{min}$ not as a sharp cutoff, but as a fundamental scale, although it regularizes the theory. Instead, it leads to exponential suppression of high-energy modes that can be described in terms of an effective dimensional reduction, or effective multifractality of spacetime -- the latter aspect is partially in common with \cite{Modesto_2009,Modesto_2010}. These features complement the minimal length paradigm and may offer an alternative route toward ultraviolet completeness in quantum field theory.

Finally, while the present work lays the conceptual foundation of non-standard QFT scenarios, several open directions remain. The formalism is expected to be compatible with gauge theories, as the absence of a sharp cutoff preserves covariance. No conceptual obstruction is anticipated in extending the framework to fermionic fields, and $\tau_{min}$ remains consistent in that context. Among others, the interpretation of the theory as an open quantum system suggests possible phenomenological implications. For example, massless neutrino oscillations may happen in an open quantum system \cite{Benatti_2001} (in this context, also the possibility of neutrino decay has been recently proposed \cite{Stankevich:2024xyc}), and this feature
might be shared with the scenario delineated in this work. This intriguing possibility deserves further exploration.

\section*{Acknowledgments}

The research of AM is partially financed by INFN. JCV specially thanks Marco Serone, Paolo Creminelli, and Andrea Romanino for their support.

\appendix

\section{Inner Products} \label{appedix:inner_products}

Here we summarize the definition of the inner product for QMs, Nambu's generalization of QMs, wave functional formulation of QFT, and Nambu's inpired generalization of QFT provided in this paper. 
\begin{align}
\langle \Psi_1 | \Psi_2 \rangle &= \int d^3 x\, \Psi_1[x]^{\dagger} \Psi_2[x]  
&&\text{(QM)} 
\label{eq:inner-product-QM} \\
\langle \Psi_1 | \Psi_2 \rangle &= \int d^3 x\, dt\, \Psi_1[x,t]^{\dagger} \Psi_2[x,t]  
&&\text{(Nambu's QM)} 
\label{eq:inner-product-Nambu-QM} \\
\langle \Psi_1 | \Psi_2 \rangle &= \int \mathcal{D}\phi\, \Psi_1[\phi]^{\dagger} \Psi_2[\phi]  
&&\text{(Functional QFT)} 
\label{eq:inner-product-FQFT} \\
\langle \Psi_1 | \Psi_2 \rangle &= \int \mathcal{D}\phi\, dt\, \Psi_1[\phi,t]^{\dagger} \Psi_2[\phi,t]  
&&\text{(Nambu-like Functional QFT)} 
\label{eq:inner-product-Nambu-FQFT}
\end{align}

\section{Derivation of the integral formula \eqref{rep1}}
\label{appendix:newderivation}

In this appendix, we detail the evaluation of the integral in Eq.~\eqref{rep1} to show how it exactly yields the standard Feynman propagator. We start from the integral expression:
\begin{equation}
    I = -i \int_0^{\infty} d\tau \, \tau \int_0^1 d\rho \, e^{-i\rho\lambda_p\tau - i(1-\rho)\lambda_a\tau} \,,
\end{equation}
where $\lambda_p = p_0 - \omega_{\vec{p}}$ and $\lambda_a = p_0 + \omega_{\vec{p}}$. We first rewrite the argument of the exponential by factoring out $\tau$:
\begin{equation}
    \rho\lambda_p + (1-\rho)\lambda_a = \rho(p_0 - \omega_{\vec{p}}) + (1-\rho)(p_0 + \omega_{\vec{p}}) = p_0 + \omega_{\vec{p}}(1 - 2\rho) \equiv A(\rho) \,.
\end{equation}
Thus, the integral becomes
\begin{equation}
    I = -i \int_0^1 d\rho \int_0^{\infty} d\tau \, \tau \, e^{-i A(\rho) \tau} \,.
\end{equation}
To ensure convergence of the integral at $\tau \to \infty$, we implicitly shift the frequency by a small imaginary part $A(\rho) \to A(\rho) - i\epsilon$. Using the standard integration formula $\int_0^\infty dx \, x \, e^{-\alpha x} = 1/\alpha^2$, with $\alpha = i A(\rho) + \epsilon$, the $\tau$-integration yields:
\begin{equation}
    \int_0^{\infty} d\tau \, \tau \, e^{-i A(\rho) \tau} = \frac{1}{(i A(\rho))^2} = -\frac{1}{A(\rho)^2} \,.
\end{equation}
Substituting this back into the $\rho$-integral, we get:
\begin{equation}
    I = i \int_0^1 d\rho \, \frac{1}{[p_0 + \omega_{\vec{p}}(1 - 2\rho)]^2} \,.
\end{equation}
We evaluate this integral by making the substitution $u = p_0 + \omega_{\vec{p}} - 2\rho\omega_{\vec{p}}$, which gives $du = -2\omega_{\vec{p}} d\rho$. The integration limits change from $\rho \in [0,1]$ to $u \in [\lambda_a, \lambda_p]$. The integral
becomes:
\begin{equation}
I = i \int_{\lambda_a}^{\lambda_p} \frac{1}{u^2} \left( \frac{du}{-2\omega_{\vec{p}}} \right)  = \frac{i}{2\omega_{\vec{p}}} \left( \frac{\lambda_a - \lambda_p}{\lambda_a \lambda_p} \right) \,.
\end{equation}
Recalling the definitions of $\lambda_p$ and $\lambda_a$, we evaluate the numerator and the denominator:
\begin{align}
\lambda_a - \lambda_p &= (p_0 + \omega_{\vec{p}}) - (p_0 - \omega_{\vec{p}}) = 2\omega_{\vec{p}} \,, \nonumber \\
\lambda_a \lambda_p &= (p_0 + \omega_{\vec{p}})(p_0 - \omega_{\vec{p}}) = p_0^2 - \omega_{\vec{p}}^2 = p_0^2 - \vec{p}^2 - m^2 = p^2 - m^2 \,.
\end{align}
Inserting these into our expression, the factor of $2\omega_{\vec{p}}$ cancels out:
\begin{equation}
    I = \frac{i}{2\omega_{\vec{p}}} \left( \frac{2\omega_{\vec{p}}}{p^2 - m^2} \right) = \frac{i}{p^2 - m^2 + i\epsilon} \,,
\end{equation}
which is exactly the standard Feynman propagator.

\section{On the interpretation of particle propagation in proper-time QFT} \label{appendix:interpretation}

Here we elaborate on the interpretation of the Green function in the context of Nambu-like QFT. To this end, we follow the main arguments presented in Chapter 33 of \cite{Schwartz:2014sze}, but adapted to the new approach of this work.

The Green function in configuration space can be expressed as
\begin{align}
D_F(x-y) &= \int\frac{d^4p}{(2\pi)^4} \frac{ie^{-i(x-y)}}{p^2-m^2+i\,\epsilon}
\nonumber \\
& =\int_0^\infty d\tau \int_0^{1}\, d\rho  \int \frac{d^4p}{(2\pi)^4} \langle y|p\rangle \left(-i\tau\right) e^{-i\rho\lambda_p\tau} e^{-i(1-\rho)\lambda_a\tau}  \langle p|x\rangle ,
\end{align}
with
\begin{equation}
\langle p|x\rangle = e^{-ip\cdot x}, \quad \langle y|p\rangle = e^{+ip\cdot y} ,
\end{equation}
thus
\begin{equation}
D_F(x-y) = -i \int_0^\infty d\tau \, \tau \int \frac{d^4p}{(2\pi)^4} \langle y|p\rangle \langle p| \int_0^1 e^{-i\rho\hat{\lambda}_p\tau} e^{-iz(1-\rho)\hat{\lambda}_a\tau} d\rho \ket{x}\,,
\end{equation}
where the hat notation means operators. The energy parameters are defined as
\begin{align}
\hat{\lambda}_p &= \hat{\omega}_k - \hat{E} \\
\hat{\lambda}_a &= \hat{E} + \hat{\omega}_k .
\end{align}
Using the completeness relation $\int \frac{d^4p}{(2\pi)^4} |p\rangle\langle p| = \mathbf{1}$, we find
\begin{equation}
D_F(x-y) = -i \int_0^\infty d\tau \, \tau \int_0^1 d\rho \langle y| e^{-i\rho\hat{\lambda}_p\tau} e^{-iz(1-\rho)\hat{\lambda}_a\tau} \ket{x} .
\end{equation}
We can introduce the proper time states,
\begin{equation}
D(x-y) = -i \int_0^\infty d\tau  \, \tau\int_0^1 d\rho \langle y|x,\tau,\rho\rangle
\end{equation}
where
\begin{equation}
|x,\tau,\rho\rangle = e^{-i\rho\hat{\lambda}_p\tau} e^{-i(1-\rho)\hat{\lambda}_a\tau} |x\rangle
\end{equation}
Alternatively, this can be written as
\begin{equation}
|\tau,x\rangle = \int_0^1 d\rho\, e^{-i\rho\hat{\lambda}_p\tau} e^{-i(1-\rho)\hat{\lambda}_a\tau} |x\rangle
\end{equation}
The Feynman propagator $D_F(x-y)$ in configuration space represents the probability amplitude for a quantum excitation to propagate from spacetime point $x$ to point $y$. Unlike its momentum-space representation, which involves a simple pole structure corresponding to on-shell particles, the configuration space form integrates over proper time $\tau$ and thereby encompasses all possible intermediate mass-shell configurations. The proper time formulation makes manifest that the propagator cannot be understood as describing a single classical trajectory, but rather emerges from a quantum superposition over world-lines of all possible proper time durations. The auxiliary parameter $\rho$  further enriches this picture by revealing that each such world-line itself represents a superposition of particle and antiparticle character, weighted by how the total proper time is apportioned between forward and backward temporal propagation.

\paragraph{Physical interpretation.}
The proper time formulation reveals a striking physical picture of particle propagation in quantum field theory. When a particle travels from point A to point B over a total proper time $\tau$, it does not propagate purely as a particle or purely as an antiparticle. Rather, the propagator integrates over all possible divisions of the proper time between particle and antiparticle states.

This interpretation emerges directly from the structure of the equation in \eqref{rep1}. The parameter $\rho \in [0,1]$ represents the fraction of proper time spent in the particle state (with eigenvalue $\hat{\lambda}_p = \hat{\omega}_{\vec{p}} - \hat{E}$), while the remaining fraction $(1-\rho)$ is spent in the antiparticle state (with eigenvalue $\hat{\lambda}_a = \hat{E} + \hat{\omega}_{\vec{p}}$). The exponential factors $e^{-i\rho\hat{\lambda}_p\tau}$ and $e^{-i(1-\rho)\hat{\lambda}_a\tau}$ describe the phase evolution during these respective portions of the journey.

The propagator sums over all possible proper time divisions: from $\rho=0$ (purely antiparticle propagation, or equivalently, backward-in-time particle propagation) to $\rho=1$ (purely forward-in-time particle propagation), with all intermediate superpositions contributing. This provides a concrete realization of the particle-antiparticle duality inherent in relativistic quantum field theory.

\section{Renormalization Group with a Proper Time}\label{appendix:RGE}

In the proper-time Schrödinger formulation, ultraviolet divergences are naturally regulated by the auxiliary parameter $\tau$. After Wick rotation to Euclidean space, the Schwinger representation reads:
\begin{equation}
\frac{1}{p_E^2 + m^2} = \int_0^\infty ds \, e^{-s (p_E^2 + m^2)}.
\end{equation}
For our purpose, it is convenient and sufficient to work here with the Schwinger $s$–representation in Euclidean space, which is equivalent to the $\tau$-based representation discussed above.

Introducing a UV regulator $s$:
\begin{equation}
G_E(p;s) = \int_{s}^\infty ds' \, e^{-s' (p_E^2 + m^2)} = \frac{e^{-s (p_E^2 + m^2)}}{p_E^2 + m^2}.
\end{equation}
The regulator $s$ corresponds to a regulator $\tau$ in the $\tau$-representation \eqref{rep1}, such that $s=\tau^2$ (so at the level of the lower integration limit only).
The exponential factor implements a covariant UV damping, rendering loop integrals finite and suggesting an effective dimensional reduction at high energies.
 
Therefore, the proper time can play the role of the RG scale:
\begin{align}
& s \frac{d}{ds} \alpha(s) = -\frac{1}{2}\,\beta(\alpha)  \nonumber \\
& \tau \frac{d}{d\tau} \alpha(\tau) = -\beta(\alpha),
\end{align}
where $\alpha(s)$ ($\alpha(\tau)$) is the running coupling defined at the scale $s$ ($\tau$). This formulation provides an alternative interpretation of renormalization, linking the RG flow to the dynamical evolution in proper time. For a recent approach to RG in terms of the Schwinger propagator, see \cite{Abel:2023ieo,Bonanno:2025tfj,Giacometti:2025qyy}. 

It is worth emphasizing that the scalar field theory discussed here serves only as an illustrative example. The proper-time Schrödinger representation is general and applies to quantum field theories beyond the scalar case. Moreover, in this section, the proper time $\tau$ is still treated as a regulator, not as a physical parameter. Loop computations are performed using standard techniques, with $\tau$ providing a smooth, Lorentz-invariant cutoff. In later sections, $\tau_{\min}$ is promoted to a physical scale, leading to structural modifications of the theory and a reinterpretation of the constrained dynamics.

\bibliographystyle{jhep}
\bibliography{biblio}

\end{document}